\documentclass[twocolumn,prl]{revtex4}

\usepackage{ORI_Group_style}

\begin{document}

\title{Strong Single-Photon Coupling in Superconducting Quantum Magnetomechanics}
  
\author{Guillem Via$^{1,2}$}
\author{Gerhard Kirchmair$^{1,3}$}
\author{Oriol Romero-Isart$^{1,2}$}

\affiliation{$^{1}$Institute for Quantum Optics and Quantum Information of the
Austrian Academy of Sciences, A-6020 Innsbruck, Austria.}
\affiliation{$^{2}$Institute for Theoretical Physics, University of Innsbruck, A-6020 Innsbruck, Austria.}
\affiliation{$^{3}$Institute for Experimental Physics, University of Innsbruck, A-6020 Innsbruck, Austria.}

\begin{abstract}
We show that the inductive coupling between the quantum mechanical motion of a superconducting microcantilever and a flux-dependent microwave quantum circuit can attain the strong single-photon nanomechanical coupling regime with feasible experimental parameters. We propose to use a superconducting strip, which is in the Meissner state, at the tip of a cantilever.  A pick-up coil collects the flux generated by the sheet currents induced by an external quadrupole magnetic field centered at the strip location. The position-dependent magnetic response of the superconducting strip, enhanced by both diamagnetism and demagnetizing effects, leads to a strong magnetomechanical coupling to quantum circuits. 
\end{abstract}

\maketitle

\begin{figure}[t]
\centering
\includegraphics[width= \columnwidth]{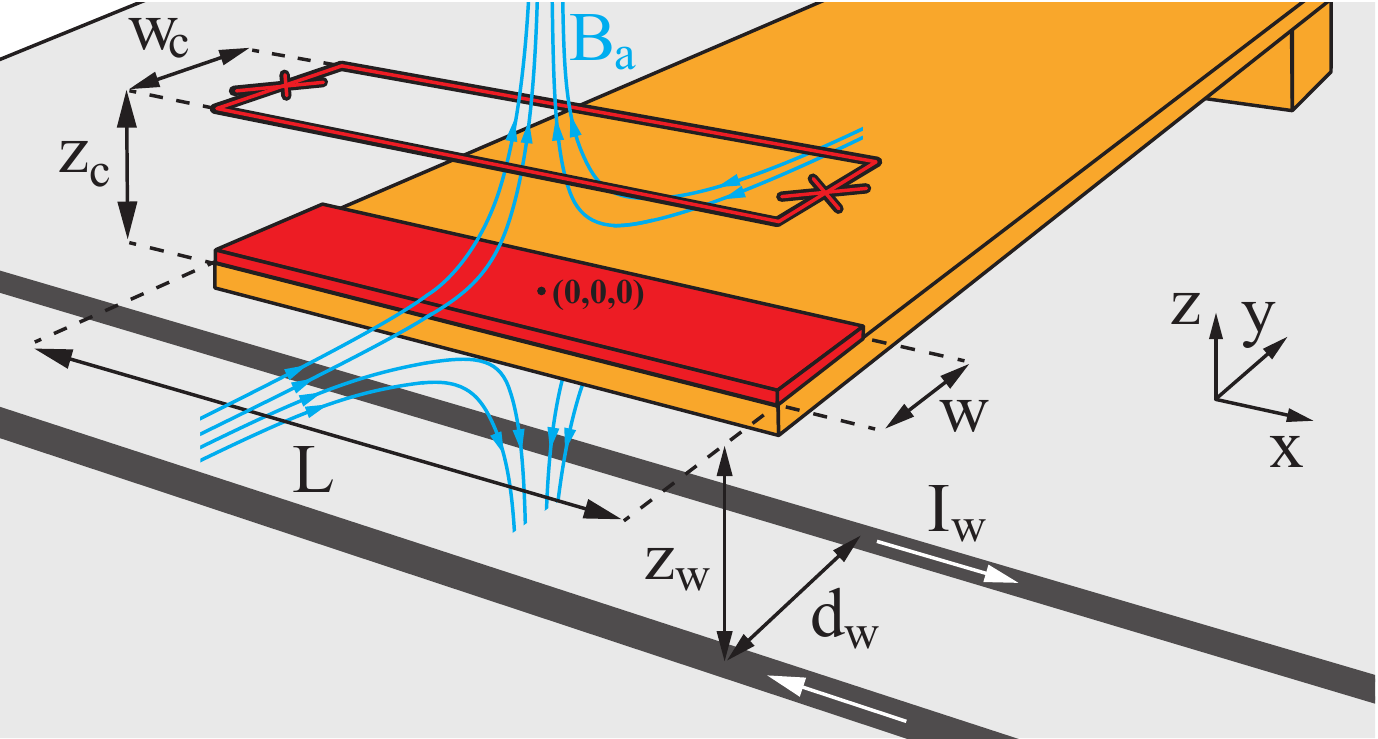}
\caption{ (Color online) Schematic illustration of the proposal (not to scale). A superconducting strip of length $L$ and width $w$ is deposited on the tip of the cantilever. At a distance $z_c$ above the cantilever a pick-up coil of the same length and width $w_c$, which is fabricated on a wafer not shown for clarity, collects the flux generated by the currents in the strip induced by an external quadrupole field $\BB_a$. The $\BB$-field is generated by two parallel wires with opposite current of intensity $I_w$, separated by a distance $d_w$, and placed below the cantilever at a distance $z_w$. An additional perpendicular bias field creates a zero-field at the strip position.}
\label{Fig1}
\end{figure}

In quantum nanomechanics, the strength of the radiation-pressure interaction between a single electromagnetic mode of frequency $\w$ and a micromechanical mode of frequency $\Omega$ and effective mass $M$ is denoted by $g_0$, the so-called single-photon coupling rate~\cite{ReviewOpto}. This is the cavity frequency shift due to a zero-point motion displacement of the mechanical oscillator, given by $z_{zp} = [\hbar/(2 M \Omega)]^{1/2}$, namely $g_0 = z_{zp} \partial \w/\partial z|_{z=0}$. The single-photon coupling, being non-linear, could be exploited to observe non-Gaussian physics in micromechanical oscillators~\cite{Mancini1997,Bose1997,Ludwig2008,Rabl2011,Nunnenkamp2011,Qian2012}, a goal that would represent a milestone in the field~\cite{ReviewOpto}. However this is today experimentally very challenging. The mechanical mode (electromagnetic mode) suffers decoherence with a rate $\Gamma$ ($\kappa$) whose origin depends on the particular experimental implementation. To fully exploit the non-Gaussian character of the single-photon nanomechanical coupling one would like to operate in the strong-coupling regime $g_0 \gtrsim \Gamma, \kappa$ as well as in the resolved sideband regime $\Omega /\kappa \gtrsim 1$. The latter is required to sideband cool the mechanical mode into the ground state~\cite{Genes2008,Wilson-Rae2007,Marquardt2007}. While $g_0/\Gamma \gtrsim 1$ and $\Omega/\kappa \gtrsim 1$ has been achieved simultaneously in several experiments~\cite{ReviewOpto}, the so-called single-photon strong coupling regime $g_0/\kappa \gtrsim 1$ is much more challenging. Indeed, according to \cite{ReviewOpto}, the highest values of $g_0/\kappa$ obtained so far with solid mesoscopic objects are $\sim 10^{-3}$~\cite{Chan2011,Palomaki2013} (with cold gases one achieves $g_0/ \kappa \sim 1$~\cite{Murch2008,Brennecke2008} but not in the resolved sideband regime). 

In this Letter we propose a microwave optomechanical scenario, see \figref{Fig1}, where we show that the strong single-photon regime $g_0/\kappa \gtrsim 1$ can be achieved in the resolved sideband regime with feasible experimental parameters.
Contrary to most of the current experiments in microwave optomechanics~\cite{ReviewOpto,Palomaki2013}, where the optomechanical coupling is implemented capacitively, here we motivate to use an inductive coupling to a flux-dependent quantum circuit as a way to obtain three orders of magnitude stronger couplings. Such a strong quantum {\em magnetomechanical} (MM) coupling is achieved via the magnetic response of a superconducting (SC) strip in an inhomogenous external field that is strengthened by the large diamagnetic and demagnetizing effects of superconducting strips in the Meissner state~\cite{BrandtIndenbom,ZeldovClem}. This contrasts to other experiments and proposals on quantum magnetomechanics that do not exploit this fact and thus do no achieve such strong couplings, see for instance~\cite{Etaki2008,Nation2008,Lambert2008,Xia2009,Xia2010}.

The quantum MM coupling to a flux-dependent quantum circuit can be obtained as follows. While in principle one just requires a quantum circuit with a SQUID loop, here we use the particular example of a transmon qubit~\cite{Koch2007} that operates as a slightly anharmonic LC oscillator with creation (annihilation) mode operators $\adop$ ($\aop$). The Hamiltonian can be written as $\Hop = \hbar \omega (z_m) \adop \aop + \hbar \beta \adop \adop \aop \aop/2$, where $\hbar \omega (z_m)  = [8 E_J(z_m) E_C]^{1/2}- E_C$,  $E_J(z_m) = 2 E_{J_1} \cos \spare { \pi \Phi (z_m)/\Phi_0 }$, $\hbar \beta = - E_C$, $\Phi_0$ is the flux quantum, and $z_m$ is the position of the superconducting strip along the $z$-axis, as described in more detailed below. Here $E_C$ is the charging energy of a single electron stored in the capacitance, and $E_{J_1}$ is the energy associated with an electron tunneling across one of the two identical junctions. The transmon regime requires $E_J / E_C \gg 40$. Hereafter we will not use the anharmonic term, which can be a resource for many applications, and will only focus on the flux-dependent microwave harmonic oscillator. The flux threading the pick-up coil $\Phi(z_m)$ depends on the $z$-displacement of the mechanical oscillator from the equilibrium position $z_m=0$, which is given by $z_m = z_{zp} (\bdop+\bop)$. By expanding $\omega (z_m)$ around $z_m=0$ one arrives at the standard single-photon coupling nanomechanical Hamiltonian~\cite{ReviewOpto} $\hat H = \hbar \w \adop \aop + \hbar \Omega \bdop \bop  - \hbar g_0 \adop\aop (\bdop + \bop)$ with $\w=\w(0)$ and $g_0 = \phi \omega_0 \eta$, where $\hbar \omega_0 \equiv [8 E_{J_1} E_C]^{1/2}$, $\phi \equiv  \pi \sin \pare { \pi \Phi (0) / \Phi_0}/ [2 \cos \pare { \pi \Phi (0)/\Phi_0}]^{1/2} $, and
\be \label{eta}
\eta \equiv \frac{z_{zp}}{\Phi_0} \left. \fpd{\Phi(z)}{z} \right|_{z=0}.
\ee
The dimensionless parameter $\eta$ quantifies the MM coupling to any quantum circuit since it is the variation of flux (in units of $\Phi_0$) in the pick-up coil due to a zero-point motion displacement of the mechanical oscillator. The decoherence rate of the quantum circuit can be generally expressed as $\kappa = \omega_0/Q$, where $Q$ is the circuit quality factor. Therefore, the ratio between the single-photon coupling and $\kappa$ is given by $g_0/\kappa = \phi Q \eta$. The parameter $\phi$ can be tuned by varying $\Phi (0)$. Consequently the MM coupling can be switched-on (switched-off) by operating at the linear (quadratic) regime, \eg~$\Phi(0)/\Phi_0 \sim 1/4 $ (\eg~$\Phi(0)/\Phi_0 = 0 $), where $\phi \sim 2$ ($\phi = 0$).
Note that since values of $Q \sim 10^6$ have been measured~\cite{Paik2011,Rigetti2012}, the strong single-photon regime $g_0/\kappa \gtrsim 1$ could be thus achieved provided $2 \eta \gtrsim 10^{-6}$. In the following we propose and analyze a setup where such regime could be achieved. 

We consider a thin SC strip occupying the region $x \in [-L/2, +L/2]$, $y \in [-w/2,+w/2]$ and $z \in [z_m-t/2,z_m+t/2]$, with $L \gg w \gg t$, see Fig. \ref{Fig1}. The SC strip is assumed to harmonically oscillate along the $z$-axis, with equilibrium position at $z_m=0$ and harmonic frequency $\Omega$. This can be achieved, for instance, by depositing the SC strip at the tip of a non-magnetic micromechanical cantilever of thickness $t_0$, width $L$, and mass density $\rho_0$, see \figref{Fig1}. In the calculation of the single-photon radiation pressure coupling, the effective mass of the mechanical oscillator can be approximated by~\cite{Pinard1999} $M=L w (\rho t + \rho_0 t_0)$, where $\rho$ is the mass density of the SC material. A rectangular pick-up coil covering the area $x \in [-L/2, +L/2]$, $y \in [-w_c/2,+w_c/2]$ is placed at $z=z_c$ on a second wafer.
The SC strip, which is considered to be in the Meissner state, fulfills that either the London penetration depth $\lambda \ll t$ or, if $\lambda \gtrsim t$, the two-dimensional screening length $\Lambda \equiv \lambda^2/t \ll w$~\cite{BrandtIndenbom, ZeldovClem}. It is also assumed that $t >  \xi$, where $\xi$ is the superconducting coherence length. Under these standard conditions one can treat the magnetic response of the SC strip using London theory~\cite{BrandtIndenbom, ZeldovClem}.

The MM coupling is established by applying an external $\BB$-field to the SC strip in the Meissner state. Due to the diamagnetic response of the SC strip, currents are induced to have a zero total $\BB$-field in the interior of the sample~\cite{BrandtIndenbom, ZeldovClem}. The flux threading the pick-up coil generated by the induced strip currents depends on the strip $z$-position of the cantilever. Stronger couplings are obtained when an inhomogeneous field with a gradient along $z$ is applied. The reason is that the induced currents depend in this case on the position of the cantilever and therefore $\eta$ scales as $1/z_c$ for $z_c \gtrsim w$. This contrasts with the case of a homogenous applied field since then the position-dependent flux only arises because the distance between the cantilever and the pick-up coil changes, thereby leading to $\eta \propto 1/z_c^2$. A convenient inhomogeneous magnetic field, with a gradient along $z$, and uniform along the $x$-axis (the long axis of the strip), is given by the quadrupolar field 
$\mathbf{B}_a(y,z)=b(- y \uey+ z \uez)$, where the gradient $b$ is constant and its maximum value is limited to ensure field strengths in the strip are below its critical field.

The induced currents in the SC strip in the presence of the applied field $\BB_a$ can be calculated as follows. Since one needs the field generated by the induced currents at a distance $z_c \gg t$, one can use the average sheet current $ \mathbf{K}(y,z_m) \equiv \int_{z_m-t/2}^{z_m+t/2}{\mathbf{J}(y,z)\textrm{d}z}$, where $\mathbf{J}$ is the volume current density. The currents are assumed to be independent on $x$ since the applied field is homogeneous in $x$, $L \gg w$, and when  $t \ll w$ the current distribution is not affected by the $y$ component of the external field~\cite{Prigozhin10}. Moreover, we show in the supplementary material (SM) that although $\mathbf{B}_a$ is not uniform across $t$, the induced $K_x$ only depends on the thickness-averaged external vector potential under the thin film approximation. Hence, when the strip is at some height $z=z_m$, $K_x$ will be well approximated to that induced by a uniform out-of-plane field $\mathbf{B}_a = b z_m \mathbf{\hat{z}}$, namely~\cite{BrandtIndenbom, ZeldovClem} 
\begin{equation} 
\mathbf{K}(y,z_m)=\frac{b z_m}{\mu_0} \frac{2 y}{\sqrt{\left(w/2\right)^2-y^2}} \uex,
\label{thinK}
\end{equation} 
where $\mu_0$ is the vacuum permeability. 
This current distribution expels the out-of-plane $\BB$-field from the interior of the sample, depends on the position of the cantilever $z_m$, and is zero when $z_m=0$. 

\begin{figure}
\centering
\includegraphics[width=\columnwidth]{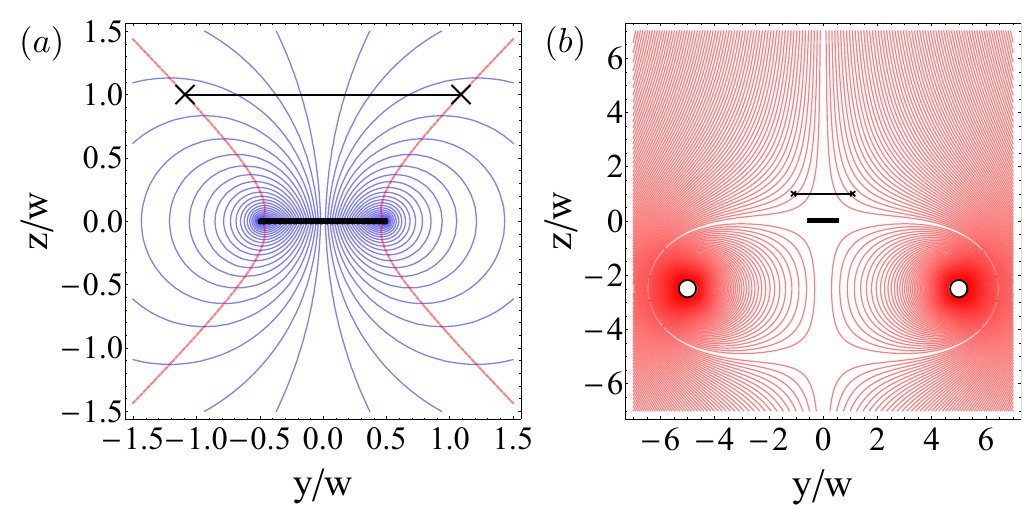}
\caption{ (Color online) $\BB$-field lines corresponding to the (a) field created by the induced currents $\BB_K$ and (b) the applied field created by two antiparallel wires (marked as white circles) and a bias field, see text. The SC strip (pick-up coil) is illustrated to scale in both plots with a solid rectangle (solid line segment delimited by crosses). The dashed red line in (a) marks the optimal pick-up coil width $w_c^\star$ for each pick-up coil height $z_c$. }
\label{Fig2}
\end{figure}

\begin{figure*}
\centering
\includegraphics[width=\linewidth]{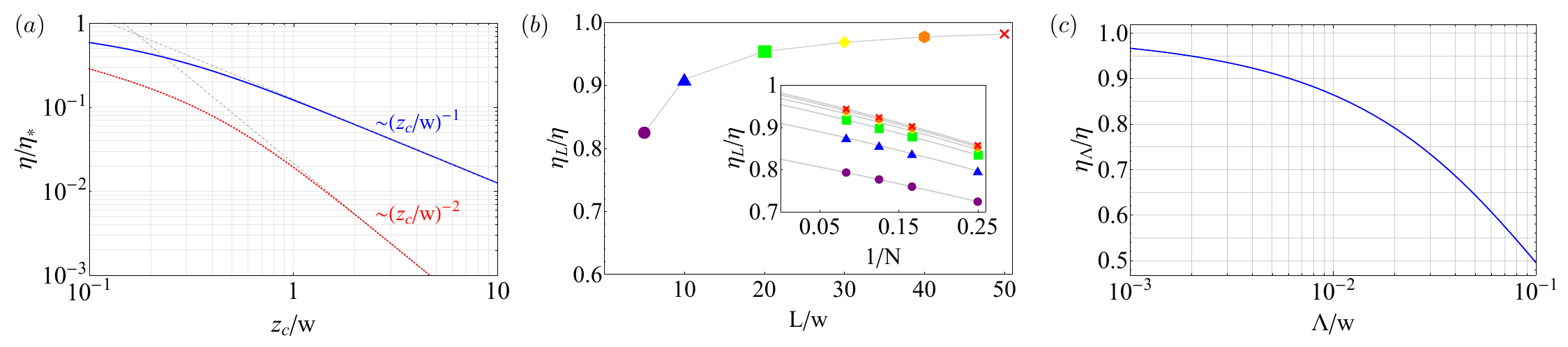}
\caption{ (Color online)  (a) $\eta/\eta_\star$ as a function of $z_c/w$ for the case of the quadrupole field (red blue line) corresponding to \eqcite{chi} and for a homogeneous field (dotted red line) corresponding to the expression given in the SM. Dashed grey lines indicate the asymptotic scaling for $z_c/w \gtrsim 1$. (b) $\eta_L/\eta$ values computed with MEM~\cite{Sanchez01, Chen08, Guillem13} for different finite strip lengths $L/w$. Inset shows the values obtained by extrapolating the results for different number of cells $N$ in the MEM method. (c) $\eta_\Lambda/\eta$ as a function of $\Lambda/w$ calculated using the expression for the sheet currents given in \cite{Plourde} (see also SM). 
}
\label{Fig3}
\end{figure*}

To obtain an analytical expression for $\eta$ one needs the vector potential generated by the strip currents. This can be calculated by integrating across the strip width the contributions from the infinitesimal narrow straight filaments that compose it, namely $\AB_K (y,z) = \int_{-w/2}^{w/2} \text{d} \textbf{A}_K (y,y',z)  $, where, using  Amp\`ere's law,
\begin{equation}
 \text{d} \textbf{A}_K (y,y',z) =-\frac{\mu_0 \text{d} I}{2 \pi} \ln{\spare{(y-y')^2+(z-z_m)^2}} \hat{\textbf{x}},
\label{Awire}
\end{equation} 
with $\text{d} I =  K_x(y',z_m) \text{d} y'$. Using \eqnref{thinK} one obtains that $\mathbf{A}_K (y,z) = A_K (y,z) \uex$ is given by
\begin{equation}
\begin{split}
\frac{A_K(y,z)}{b z_m }= 
y  - \frac{y}{\abs{y}} \text{Re} \cpare{ \sqrt{\spare{y+ \textrm{i}(z-z_m)}^2 - \left(\frac{w}{2}\right)^2} }.
\label{analyticAx}
\end{split}
\end{equation}
In this particular longitudinal geometry, one can use the contour lines of the vector potential to plot the magnetic $\BB$-field lines~\cite{Brandt00} of $\BB_K = \curl \AB_K$, as shown in Fig.~\ref{Fig2}a. The magnetic flux threading the pick-up coil is given by the contour integral of the vector potential along the coil wire. This leads to $\Phi(z_m) = 2 L_c A_K(w_c/2,z_c)$. Using \eqnref{analyticAx} and recalling \eqnref{eta}, one obtains $\eta  =  \eta_\star \chi$, where
\be
 \chi \equiv \frac{w_c}{w}-  \Re \cpare{ \sqrt{ \pare{ \frac{w_c}{ w} +  \im \frac{2 z_{c}}{w} }^2 - 1} }.
 \label{chi}
\ee
The maximum value of $\eta$ is given by $\eta_\star \equiv z_{zp} b L_c w/\Phi_0$, which corresponds to the limit $w_c \rightarrow w$ and $z_{c} \rightarrow 0$. Given a coil distance $z_c$, the value of $\eta$ is maximized for an optimal $w_c^\star$ which corresponds to the width for which the lateral long wires of the pick-up coil coincide with the lines of $B^z_K=0$ (see Fig. \ref{Fig2}a). Using $w_c^\star$, $\eta/\eta_\star$ can be plotted as a function of $z_c/w$, see Fig.~\ref{Fig3}a.  At an experimentally feasible distance $z_c=w$, $\eta/ \eta_\star \approx 1.2 \times 10^{-1}  $. At the same distance, an homogeneous external field would lead to $\eta/ \eta_\star \approx 1.9 \times 10^{-2} $, nearly an order of magnitude less, see SM and \figref{Fig3}a.

The value of $\eta_\star$, and thus of $g_0/\kappa$,  is maximized when the maximum gradient $b^\text{max}$ allowing for superconductivity in the strip is used. That is, one requires  $\abs{ \mathbf{B}_a + \mathbf{B}_K} < B_c$ at any point in the sample, where $B_c$ is the first critical field from the SC strip material. By taking into account the demagnetizing effects~\cite{BrandtIndenbom, ZeldovClem}, it is shown in the SM that this leads to $b^\text{max} = f(t/w) 2 B_c/w$, where $f(x)= [1+ (\sqrt{2 x} +x) (1+x)]^{-1/2}$. Taking $L_c = L$, one arrives at
\begin{equation}
\eta_\star=   \frac{2 B_c}{\Phi_0} f(t/w) \sqrt{\frac{\rho t}{\rho t + \rho_0 t_0} \frac{\hbar}{2 \rho \Omega}}  \sqrt{ \frac{L_c}{t w} },
\label{eta*}
\end{equation}
that together with \eqnref{chi} gives an analytical expression for $\eta$ and thus $g_0/\kappa$.
Hereafter we consider niobium for the SC strip, with $B_c \approx 140 \text{ mT}$ and $\rho=8.57 \times 10^{3} \text{ kg/m$^3$}$, the strip dimensions $t = 50 \text{ nm}$, $w = 1 \text{ $\mu$m}$ and $L_c = 100 \text{ $\mu$m}$, the cantilever to be made of silica with $\rho_0= 2.3 \times 10^{3} \text{ kg/m$^3$}$, $t_0 = 0.5  \text{ $\mu$m}$, and $\Omega = 2 \pi \times 10^{6}$ Hz. Using $z_c = w$, $w_c = w^\star_c \approx 2.2 \text{ $\mu$m}$, and $b^\text{max} \approx 2.4 \times 10^5 \text{ T}/\text{m}$, one obtains $2 \eta \sim 20.4 \times 10^{-6}$. This is the main result of the Letter because $Q \approx 10^{6}$ has been experimentally measured~\cite{Paik2011,Rigetti2012}, and thus, using the maximum gradient $b_\text{max}$, this would lead to $g_0/\kappa \approx 20.4 $, well within the single-photon strong coupling regime. Mechanical dampings of $\gamma \sim 2\pi \times 1 \text{ Hz}$ have been measured in low frequency mass-loaded cantilevers \cite{Mamin2003,Rugar2004,Vinante2014}. This would lead to mechanical decoherence rates of $\Gamma \approx \gamma K_b T/(\hbar \Omega) \sim 2\pi \times 6.5  \text{ kHz}$ at $T=50 \text{ mK}$, and hence to a single-photon cooperativity $\mathcal{C} = g_0^2/(\kappa \Gamma) \sim 400$ (using the maximum gradient $b^\text{max}$).

Let us now discuss two approximations that were used to calculate $\eta$:  the distributions of fields and currents were those of (i) an infinite strip length with (ii) $\Lambda = 0$. Regarding (i), we have numerically computed  $\eta$ for $z_c=w$ and $w_c = w_c^\star$ for finite $L/w$ values. In \figref{Fig3}b the ratio of $\eta_L$ (computed with a finite $L$) with $\eta$ (obtained with an infinite length) as a function of $L/w$ is plotted. This has been done using the Magnetic Energy Minimization (MEM) method \cite{Sanchez01, Chen08, Guillem13}.  As expected $\eta_L/\eta$ approaches unity as one increases $L/w$, with $\eta_L/\eta \sim 0.98$ already at $L/w=50$ (with the values discussed above one has $L/w=100$). Regarding (ii), a finite $\Lambda$ can be taken into account by using an approximated expression for the current distribution, see~\cite{Plourde} and SM. Using this, $\eta_\Lambda/\eta$ (with $\eta_\Lambda$ being the value of $\eta$ for a finite $\Lambda$) can be plotted as a function of $\Lambda/w$, see \figref{Fig3}c. For niobium, $\lambda=39$ nm and therefore $\Lambda/w = 3 \times 10^{-2}$. This leads to $\eta_\Lambda/\eta \approx 0.73$ at $z_c=w$ and $w_c = w_c^\star$. This validates approximations (i) and (ii).

To generate the ideal quadrupole $\BB$-field given by $\BB_a = b (-y \uey + z \uez)$ we propose to use two thin long straight wires placed along the $x$-axis, at some height $z=-z_w$ and $y = \pm d_w/2$, with $d_w = 4 z_w$, see \figref{Fig2}b. The wire at $y=d_w/2$ ($y=-d_w/2$) has a positive (negative) current $I_w$, namely along $\uex$ ($-\uex$), see \figref{Fig1}.
An expression for the $\BB$-field generated by the wires $\mathbf{B}_\text{w}$ can be straightforwardly obtained, see SM.  To have a zero field at the position of the strip, namely at $z=0$ and $y=0$, one should add an out-of-plane bias field $\mathbf{B}_b = 4\mu_0 I_w/(5 \pi z_w ) \uez$. The total field $\mathbf{B}_\text{w} + \mathbf{B}_b $ is very similar to the quadrupole field $\BB_a$, see \figref{Fig2}b. In particular, the gradient along $z$ is given by  $\partial_z B_z |_{ z=0} = b [1 + \alpha (2 y/z_w)^2 + \mathcal{O}(y/z_w)^4 ]$, with $b =16 \mu_0 I_w/(25 \pi z_w^2)$, and $\alpha = 72/100$. Since $\abs{y} < w/2$, one can choose $w/z_w$ to set the maximum inhomogeneity $\epsilon = \spare{\partial_z B_z |_{y=w/2, z=0}- \partial_z B_z |_{ y=0, z=0}}/b \ll 1$ to be as small as desired by using  $w/z_w < \sqrt{\epsilon/\alpha} $. Restricting the maximum current intensity to the experimentally feasible value of $I_w=1 \text{ A}$, one has that for $z_w = 5.4 \text{ $\mu$m}$ the gradient is $b \approx 4.1 \times 10^4 \text{ T/m} \approx 0.17 \, b^\text{max} $ and therefore $g_0/\kappa \approx 3.5$ and $\mathcal C \sim 12 $, still well within the single-photon coupling regime. We have numerically validated that the inhomogeneity in the gradient field leads to negligible corrections. With this configuration the total $\BB$-field at the wire of the pick-up coil at $z_c= 1 \text{ $\mu$m}$ and $y=w_c^\star/2=1.1 \text{ $\mu$m} $ is $\sim 62 \text{ mT}$.

The intensity in the wires and the strength of the bias field might fluctuate as $I_w(t)=I_w + \delta I_w(t)$ and $\BB_b(t)=(B_b+\delta B_b(t)) \uez$. The fluctuations of the intensity (bias field) are characterized by a power spectrum $S_I(\w)$ ($S_B(\w)$), where $S_f(\w) \equiv 2 \int_0^\infty \avg{\delta f(t) \delta f(0)} \cos(\w t) \text{d} t$. Consequently, the flux threading the pick-up coil will also fluctuate as $\Phi(t) = \Phi + \delta \Phi(t)$. It is shown in the SM that $S_\Phi (\w)/\Phi_0^2 = a^2_I S_I (\w)/I_w^2 + a_b^2 S_B (\w)/B_b^2 $, where the noise amplification dimensionless parameters are $a_I=6.4 \times 10^3$ and $a_B = 1.4 \times 10^4$ (their exact expression is given in the SM). To reduce the flux noise one should thus use persistent currents and gradiometric configurations. The fluctuations on the external field might also lead to decoherence in the mechanical oscillator. As shown in the SM, the magnetic Lorentz force $\FF = \int_V \JJ(\rr) \times \BB_\text{a}(\rr) \dr$ for the external quadrupole trap leads to $\FF = - M \Omega_m^2 z_m \uez/2$, where  $\Omega_m = b w [L \pi / (4 M \mu_0)]^{1/2} = 2 \pi \times 59$ kHz $< \Omega$~\footnote{To ease the notation the harmonic frequency is calculated considering the effective mass used in the single-photon coupling. One should use the much larger effective mass relating the harmonic frequency with the stiffness coefficient. The value calculated is thus an upper bound which is shown to be already smaller that the natural frequency of the cantilever.}. Since the gradient fluctuates due to the wire intensity fluctuations, so does $\Omega_m$. As shown in \cite{Gehm1998} and in the SM, this leads to Fock state transitions from level $n$ to $n  \pm 2$ with a rate given by $R_{0 \rightarrow 2} = \pi \Omega^2 S_I(2 \Omega)/(4 I_w^2) \sim 2 \pi \times 0.5 \text{ kHz}$ for $[S_I(2 \Omega)]^{1/2}/(I_w) = 10^{-5}/\text{Hz}^{1/2}$. This is two orders of magnitude smaller than $g_0$ and therefore should not compromise the strong-coupling regime.

In conclusion, we have shown that a very strong inductive coupling can be achieved between a SC strip in the Meissner state and a flux-dependent quantum circuit. This might allow to attain the so-far experimentally  challenging single-photon coupling regime in quantum nanomechanics. Such a strong coupling could also be used to exploit a linearized nanomechanical coupling to a superconducting qubit. This proposal might be employed as an experimental testbed for quantum magnetomechanics with levitated superconducting microspheres \cite{RomeroIsart2012}. An interesting further direction for research is the possibility of exploiting type-II  SC strips with controlled SC vortices to achieve even larger couplings. In this respect, this experimental scenario might offer an alternative tool to probe the rich physics of type-II superconductivity using the high-sensitivity of microcantilvers near the quantum regime.

This work is supported by the European Research Council (ERC-2013-StG 335489 QSuperMag) and the Austrian Federal Ministry of Science, Research, and Economy (BMWFW).
We are grateful to A. S\'anchez, C. Navau, M. Aspelmeyer, R. Gross, J. Hofer, J. A. Slater, M. Trupke, and W. Wieczorek for useful discussions.

\newpage

\newpage

\newpage

\section*{SUPPLEMENTARY INFORMATION}

\section{Thin film approximation}

In this section we show that the distribution of sheet current $\mathbf{K}$ over a thin flat superconducting sample ($t/w \ll 1$) only depends on the thickness-averaged vector potential. This justifies the use of Eq.~(2) in the Letter for the current distribution at the strip subjected to the $\BB$-field given by  $\BB_a(y,z) = b (-y \uey + z\uez)$. 

We start by recalling the London equation 
\be
\mathbf{A} = \mathbf{A}_a + \mathbf{A}_J = - \mu_0 \lambda^2 \mathbf{J},
\ee
where $\AB_a$ is vector potential associated to the external field and $\AB_J$ is the vector potential created by the induced currents in the superconducting sample. The vector potential associated to the external field is of the form $\mathbf{A}_a(x, y, z) = A_a(y, z) \uex$. Together with the symmetry of our system the only non-zero component of the current density is given by
\be
\begin{split}
&\mu_0 \lambda^2 J_x(y, z) = -  A_a(y, z) +\\
&+ \frac{\mu_0}{2 \pi} \int_{-\frac{w}{2}}^{\frac{w}{2}}{ \int_{z_m-\frac{t}{2}}^{z_m+\frac{t}{2}}{ \text{d} z' \text{d} y' J_x(y', z') \ln{|\mathbf{r}-\mathbf{r'}|} }} , 
\label{JLondon}
\end{split}
\ee
where $|\mathbf{r}-\mathbf{r}'| = \sqrt{(z-z')^2+(y-y')^2}$. The Biot-Savart law has been used to write $\AB_J$ as
\be
\begin{split}
A_{J,x}(y, z) =  - \frac{\mu_0}{2 \pi} \int_{-\frac{w}{2}}^{\frac{w}{2}} \text{d} y' \int_{z_m-\frac{t}{2}}^{z_m+\frac{t}{2}}   & \text{d} z'  J_x(y', z') \times \\
& \times \ln{|\mathbf{r}-\mathbf{r'}|} ,
\label{JLondon}
\end{split}
\ee
By integrating \eqnref{JLondon} across thickness and using
\bea
K_x(y, z_m) &\equiv& \int_{z_m-\frac{t}{2}}^{z_m+\frac{t}{2}}{ \text{d} z J_x(y, z)},\\  
\bar A_a(y,z_m) &\equiv& \int_{z_m-\frac{t}{2}}^{z_m+\frac{t}{2}} \text{d} z A_a(y,z),\\
\ln \abs{\rr-\rr'} &=& \ln \abs{y-y'} + \frac{1}{2} \ln \spare{1+ \pare{\frac{z-z'}{y-y'}}^2},
\eea
one arrives at
%
\be
\begin{split}
K_x (y,z_m) =& - \frac{1}{\mu_0 \lambda^2}  \overline{A}_a(y, z_m) \\
& + \frac{ t}{2 \pi  \lambda^2} \int_{-\frac{w}{2}}^{\frac{w}{2}}{  \text{d} y' K_x(y', z_m) \ln{ |y-y'| }}  \\
&+ C(y,z_m),
\end{split}
\ee
where
\be
\begin{split}
C(y,z_m) \equiv &\frac{1}{  4 \pi \lambda^2 } \int_{-\frac{w}{2}}^{\frac{w}{2}} \text{d} y' \iint_{z_m-\frac{t}{2}}^{z_m+\frac{t}{2}}  \text{d} z \text{d} z' J_x(y', z') \times \\ 
 &\times    \ln \spare{1+ \pare{\frac{z-z'}{y-y'}}^2}.         
\end{split}
\ee
As shown in \cite{VodolazovMaksimov}, the term $C(y,z_m)$ leads to negligible contribution for $t \ll w$.

Therefore, to first order on $t/w$, the distribution of $K_x$ only depends on $\overline{A}_a(y,z_m)$ and not on the particular distribution of $A_a$ across thickness. For an infinite strip whose cross-section is centered at $(y,z)=(0,z_m)$ and is subjected to the linear $\BB$-field $\BB_a(y,z) = b (-y \uey + z\uez)$, $\overline{A}_a(y, z_m) = - b y z_m t$. Note that $\overline{A}_a(y, z_m) = - b y z_m t$ is also obtained when the strip is subjected to a uniform field given by $\BB_a = b z_m \mathbf{\hat{z}}$. Thus, for both cases one obtains the same dependence of $K_x$ on $y$.

\section{Maximum  $\BB$-field gradient}

In this section we want to obtain the maximum gradient $b^\text{max}$ that can be used while allowing for superconductivity. The total field is given by $\BB_T(y,z) = \BB_a(y,z) + \BB_K(y,z)$. The maximum gradient is thus given by solving the following equation
\be
\max_{y,z \in V} \abs{\BB_T(y,z)}  = B_c,
\ee
where $V$ is the volume occupied by the strip. One can readily see that $\max_{y,z \in V} \abs{\BB_T(y,z)} = \abs{B_T(\pm w/2,\pm t/2)}$ since the applied flied is maximum at the edges and the demagnetizing effects, very large in this geometry, give the maximum field enhancement also at the edges. Note that $\mathbf{B}_a (w/2,t/2) = -b (w/2)  \uey + b (t/2)\uez$. If one neglects the currents induced by the in-plane component of the applied field, then one can show that \cite{BrandtIndenbomSM, ZeldovClemSM} 
\be
\BB_K (w/2,t/2) \approx \sqrt{\frac{w }{2 t}} \frac{bt}{2} (\uez - \uey)  .
\ee
Then, the maximum gradient is obtained by solving $\abs{\BB_a (w/2,t/2) + \BB_K (w/2,t/2)} =B_c$. This leads to
\be
b^\text{max}= \frac{2B_c}{w} f(t/w),
\ee
where
\be
f(x) \equiv \spare{1+ \pare{\sqrt{2 x} +x} \pare{1+x}}^{-1/2}.
\ee

\section{Homogeneous external field}

In this section we obtain the value of $\eta$ for the case in which the applied $\BB$-field is given by
\be
\BB_a = B_a \uez.
\ee
The response of the SC strip for such a perpendicular homogeneous field has been thoroughly analyzed in~\cite{BrandtIndenbomSM,ZeldovClemSM}. The sheet currents are given by
\begin{equation} 
\mathbf{K}(y)=\frac{B_a}{\mu_0} \frac{2 y}{\sqrt{\left(w/2\right)^2-y^2}} \uex,
\end{equation} 
which is the same expression as Eq.~[2] in the Letter with $B_a$ replacing $b z_m$. The vector potential created by the current distribution was given in Eq.~[4] in the Letter, but now reads
\begin{equation}
\begin{split}
\frac{A_K(y,z)}{B_a }= 
y  - \frac{y}{\abs{y}} \text{Re} \cpare{ \sqrt{\spare{y+ \textrm{i}(z-z_m)}^2 - \left(\frac{w}{2}\right)^2} }.
\end{split}
\end{equation}
By noting that the flux threading the pick-up coil is given by $\Phi(z_m) = 2 L_c A_K(w_c/2,z_c)$, one can readily obtain
\be 
\eta = \frac{2 B_a L_c z_{zp}}{\Phi_0} \Re \cpare{\frac{- 2 z_c/w + \im w_c/w }{\sqrt{ \pare{w_c/w + \im 2 z_c/w}^2 - 1}}}.
\ee
Analogously to the case of the quadrupole external field, $\eta$ is maximized in this case when $w_c = w_c^\star \equiv w g(z_c/w)$, where
\be
g(x) \equiv \sqrt{\frac{3+20 x^2 - 4 x \sqrt{3+16 x^2}}{3}}.
\ee
The maximum applied field is obtained when the total field, including demagnetizing effects, does not overcome $B_c$, this leads to \cite{BrandtIndenbomSM, ZeldovClemSM} $B^\text{max}_{a} = \sqrt{2 t/w} B_c$.

\section{Sheet currents for $\Lambda \neq 0$}

There is no analytical expression for the current distribution $K_x(y)$ in the SC strip for $\Lambda \neq 0$. However, as analyzed in \cite{PlourdeSM}, for a perpendicular homogeneous field $B_a$ and $\Lambda\neq 0$, a good approximation is
\begin{equation}
K_x(y) = \frac{B_a}{\mu_0} \frac{y}{\sqrt{ h_1(\Lambda/w)  \left[ \left( w/2 \right)^2 - y^2 \right] + h_2(\Lambda/w)  \Lambda w }},
\label{thinKLambda}
\end{equation}
where 
\bea
h_1(x) &=& \inv{4} - 0.63 \sqrt{x} + 1.2 x^{0.8}\\
h_2(x) &=& \half \pi + x.
\eea

Using this current distribution one can then proceed as in the manuscript and compute numerically  $\eta$, as we did in \figref{Fig3}c.

\section{$\BB$-field from the wires}

As discussed in the manuscript, to generate the quadrupole $\BB$-field given by $\BB_a(y,z) = b (-y \uey + z\uez)$ we propose to use two thin long straight wires placed along the $x$-axis, at some height $z=-z_w$ and $y = \pm d_w/2$, with $d_w = 4 z_w$. This configuration is chosen from taking only the lower coil of an anti-Helmholtz setup. The wire at $y=d_w/2$ ($y=-d_w/2$) has a positive (negative) current $I_w$, namely along $\uex$ ($-\uex$). The vector potential generated by a wire has already been given in Eq.~[3] in the Letter. One can then show that the $\BB$-field generated by the two wires is given by
\be
\mathbf{B}_w(y, z) =  \frac{\mu_0 I_w}{2 \pi z_w} \sum_{j=0}^1{ (-1)^j \BB_j(y,z) }, 
\ee
where
\be
\BB_j(y,z) \equiv  \frac{-\spare{z/z_w + 1/2}\uey + \spare{y/z_w - (-1)^j} \uez}{ \spare{y/z_w - (-1)^j }^2 + \spare{z/z_w + 1/2}^2}
\ee
One readily sees that
\be
\BB_w (0,0) = - \frac{4}{5}  \frac{\mu_0 I_w}{\pi z_w} \uez.
\ee
For this reason we propose to add a bias field $\mathbf{B}_b = - \BB_w (0,0)$ to obtain a zero field at the position of the strip, as is the case for $\BB_a(y,z) = b (-y \uey + z\uez)$.

\section{Flux noise power spectrum} \label{Fluctuations}

In this section we want to obtain the flux noise in the pick-up coil induced by the noise in the applied field. We consider the configuration in which the applied field is generated by the bias field plus the wire, whose values fluctuate as
\begin{eqnarray}
I_{w,i}(t)&=& (-1)^{i} \spare{I_w+\delta I(t)} \\
B_b(t) &=& B_b + \delta B_b(t),
\label{sourcesnoise}
\end{eqnarray}
Note that we assume that the current in the two wires is created by the same source and thus is subjected to the same noise (with the corresponding change of sign). We assume that the bias field and the intensity noise are uncorrelated, namely $\avg{\delta B_b(t) \delta I(t') }=0$ (we assume $\avg{\delta B_b(t) }= \avg{\delta I_w(t)}=0$). 
The noise will be characterized by one-side power spectrum functions defined as
\be
S_f(\w) = \frac{2}{\pi} \int_0^\infty \avg{f(t) f(0)} \cos(\w t)\text{d} t 
\ee
These bias field and intensity fluctuations will induce  flux fluctuations in the pick-up coil whose power spectrum will be given by
\be \label{eq:SPhi}
 \frac{S_\Phi(\w)}{\Phi_0^2} = a_B^2  \frac{S_B(\w)}{B^2_b} + a_I^2  \frac{S_I(\w)}{I_w^2},
\ee
where $a_{B(I)}$ are dimensionless parameters that we obtain in the following.

The flux fluctuations threading in the pick-up coil can be written as $\delta \Phi(t) = \sum_{i=1}^4 \delta \Phi_i(t)$ where $\delta \Phi_1$ ($\delta \Phi_2$) is the contribution due to the wire at $y=d_w/2$ ($y=-d_w/2$), $\delta \Phi_3$ is due to the bias field, and $\delta \Phi_4$ is due to the induced currents in the superconducting strip. Using the Biot-Savart law one can show that
\begin{eqnarray}
\delta \Phi_1 (t) &=& \delta \Phi_2 (t)= -\zeta  L_c z_w \frac{5}{16}  \frac{B_b}{I_w}  \delta I_w(t) \\
\delta \Phi_3 (t) &=& L_c w_c \delta B_b(t)\\
\delta \Phi_4 (t) &=&  \chi L_c w  \spare{\delta B_b (t) - \frac{B_b}{I_w} \delta I_w(t)}
\end{eqnarray}
where we have defined
\be
\zeta \equiv   \ln{ \left[ \frac{ \left( w_c + 4 z_w \right)^2 + 4 (z_c + z_w)^2 } { \left(w_c - 4 z_w \right)^2 + 4 (z_c + z_w)^2} \right] }. 
\ee
Recall also the definition of $\chi$ in Eq.~[5] in the Letter.
Using this result, one can readily obtain that $S_\Phi(w) = 2 \int_0^\infty \avg{\Phi(t) \Phi(0)} \cos(\w t)\text{d} t/\pi  $ can be written as \eqnref{eq:SPhi} with
\begin{eqnarray}
a_I &=& \frac{B_b L_c z_w}{\Phi_0}  \pare{  \frac{5 \zeta}{8}   +  \chi \frac{ w }{z_w} }, \nonumber \\
a_B &=& \frac{B_{b} L_c w_c}{\Phi_0} \pare{ 1  +  \chi \frac{w}{w_c}   }. \nonumber \\
\end{eqnarray}

\section{Magnetic force}

The force exerted by the external field $\mathbf{B}_a$ on the superconducting strip is given by the Lorentz force
\begin{equation}
\mathbf{F} = \int_V{ \mathbf{J} \times \mathbf{B}_a \text{d}^3 r},
\end{equation}
where $V$ is the SC sample. In our case, the currents flow along the $x$-axis direction and extend over an infinite length (we use the approximation $L \gg w$). Therefore, we the force per strip unit length, namely
\be
\frac{F_z}{L} = \int_{-w/2}^{+w/2}{\int_{z_m-t/2}^{z_m+t/2}{ J_x(y, z, z_m) B_{a,y}(y, z) \text{d} z} \text{d} y}.
\ee 
Using $\BB_a = b (-y \uey + z\uez)$ and Eq.~[2] in the Letter, one readily arrives at
\be
\frac{F_z}{L} = - \frac{\pi b^2}{\mu_0} \left( \frac{w}{2} \right)^2 z_m.
\ee
As expected, this force creates an additional harmonic potential to the motion of the cantilever. The harmonic frequency is given by
\be
\Omega_m= \frac{bw}{2} \sqrt{\frac{\pi L}{\mu_0 M }} .
\ee

As discussed in the manuscript, in the two wires configuration the gradient is given by
\be
b = \frac{16}{25} \frac{\mu_0 I_w}{ \pi z_w^2}.
\ee
As discussed in the previous section, $I_w$ fluctuates in time and thereby the gradient $b(t) = b + \delta b(t)$, where $\delta b(t) = (16/25) \mu_0 \delta I(t)/ \pi z_w^2$. Thus, also the harmonic potential will fluctuate as $M \Omega_m^2 (1 + \xi(t)) z_m^2$, where
\begin{equation}
\xi(t) \equiv \frac{ [ 2 b + \delta b (t) ] \delta b(t)}{b^2} \approx  2 \frac{\delta b(t)}{b} = 2\frac{\delta I_w(t)}{I_w} ,
\end{equation}
As discussed in \cite{Gehm1998SM}, this trap fluctuation lead to motional heating in the cantilever by inducing transitions from the ground Fock state $\ket{n=0}$ to $\ket{n=2}$ with a rate given by
\be
R_{0 \rightarrow 2} = \frac{\pi \Omega^2}{4} \frac{ S_I(2 \Omega)}{I_w^2},
\ee
where $S_I(\w)$ has been introduced in the previous section. We assume that the total frequency of the harmonic oscillator is $\Omega_T \equiv [\Omega^2+\Omega_m^2]^{1/2} \approx \Omega$.


\begin{thebibliography}{27}


\expandafter\ifx\csname natexlab\endcsname\relax\def\natexlab#1{#1}\fi
\expandafter\ifx\csname bibnamefont\endcsname\relax
  \def\bibnamefont#1{#1}\fi
\expandafter\ifx\csname bibfnamefont\endcsname\relax
  \def\bibfnamefont#1{#1}\fi
\expandafter\ifx\csname citenamefont\endcsname\relax
  \def\citenamefont#1{#1}\fi
\expandafter\ifx\csname url\endcsname\relax
  \def\url#1{\texttt{#1}}\fi
\expandafter\ifx\csname urlprefix\endcsname\relax\def\urlprefix{URL }\fi
\providecommand{\bibinfo}[2]{#2}
\providecommand{\eprint}[2][]{\url{#2}}

\bibitem{ReviewOpto}
M. Aspelmeyer, T. J. Kippenberg, and F. Marquardt, arXiv:1303.0733.

\bibitem{Mancini1997}
S. Mancini, V. I. Man'ko, and P. Tombesi, \PRA{55}{3042}{1997}.
\bibitem{Bose1997}
S. Bose, K. Jacobs, and P. L. Knight, \PRA{56}{4175}{1997}.
\bibitem{Ludwig2008}
M. Ludwig, B. Kubala, and F. Marquardt, \NJP{10}{95013}{2008}.
\bibitem{Rabl2011}
P. Rabl, \PRL{107}{063601}{2011}.
\bibitem{Nunnenkamp2011}
A. Nunnenkamp, K. Borkje, and S. M. Girvin, \PRL{107}{063602}{2011}.
\bibitem{Qian2012}
J. Qian, A. A. Clerk, K. Hammerer, and F. Marquardt, \PRL{109}{253601}{2012}.

\bibitem{Genes2008}
C. Genes, D. Vitali, P. Tombesi, S. Cigan, and M. Aspelmeyer, \PRA{77}{33804}{2008}.
\bibitem{Wilson-Rae2007}
I. Wilson-Rae, N. Nooshi, W. Zwerger, and T. J. Kippenberg, \PRL{99}{93901}{2007}.
\bibitem{Marquardt2007}
F. Marquardt, J. P. Chen, A. A. Clerk, and S. M. Girvin, \PRL{99}{93902}{2007}.

\bibitem{Chan2011}
J. Chan, T. P. M. Alegre, A. H. Safavi-Haeini, J. T. Hill, A. Krause, S. Gr\"oblacher, M. Aspelmeyer, and O. Painter, \NAT{478}{89}{2011}.

\bibitem{Palomaki2013}
T. A. Palomaki, J. D. Teufel, R. W. Simmonds, and K. W. Lehnert, \SCI{342}{710}{2013}.

\bibitem{Murch2008}
K. W. Murch, K. L. Moore, S. Gupta, and M. D. Stamper-Kurn, \NATP{4}{561}{2008}.
\bibitem{Brennecke2008}
F. Brennecke, S. Ritter, T. Donner, and T. Esslinger, \SCI{322}{235}{2008}.

\bibitem{BrandtIndenbom}
E. H. Brandt, and M. Indenbom,  \PRB{48}{12893}{1993}.

\bibitem{ZeldovClem}
E. Zeldov, J. R. Clem, M. McElfresh, and M. Darwin,  \PRB{49}{9802}{1994}.

\bibitem{Etaki2008}
S. Etaki, M. Poot, I. Mahboob, K. Onomitsu, H. Yamaguchi, and H. S. J. Van der Zant, \NATP{4}{785}{2008}.
\bibitem{Nation2008}
P. D. Nation, M. P. Blencowe, and E. Buks, \PRB{78}{104516}{2008}.
\bibitem{Lambert2008}
N. Lambert, I. Mahboob, M. Pioro-Ladri\`ere, Y. Tokura, S. Tarucha, and Y. Yamaguchi, \PRL{100}{136802}{2008}
\bibitem{Xia2009}
K. Xia and J. Evers, \PRL{103}{227203}{2009}.
\bibitem{Xia2010}
K. Xia and J. Evers, \PRB{82}{184532}{2010}.


\bibitem{Koch2007}
J. Koch, T. M. Yu, J. Gambetta, A. A. Houck, D. I. Schuster, J. Majer, A. Blais, M. H. Devoret, S. M. Girvin, and R. J. Schoelkopf, \PRA{76}{042319}{2007}.

\bibitem{Paik2011}
H. Paik, D. I. Schuster, L. S. Bishop, G. Kirchmair, G. Catelani, A. P. Sears, B. R. Johnson, M. J. Reagor, L. Frunzio, L. I. Glazman, S. M. Girvin, M. H. Devoret, and R. J. Schoelkopf, \PRL{107}{240501}{2011}.

\bibitem{Rigetti2012}
C. Rigetti, J. M. Gambetta, S. Poletto, B. L. T. Plourde, J. M. Chow, A. D. C\'orcoles, J. A. Smolin, S. T. Merkel, J. R. Rozen, G. A. Keefe, M. B. Rothwell, M. B. Ketchen, and M. Steffen, \PRB{86}{100506(R)}{2012}.

\bibitem{Pinard1999}
M. Pinard, Y. Hadjar, and A. Heidmann, Eur. Phys. J. D. {\bf 7}, 107 (1999).

\bibitem{Mamin2003}
H. J. Mamin, R. Budakian, B. W. Chui, and D. Rugar, \PRL{91}{207604}{2003}.
\bibitem{Rugar2004}
D. Rugar, R. Budakian, H. J. Mamin, and B. W. Chui, \NAT{430}{329}{2004}.
\bibitem{Vinante2014}
A. Vinante, \APL{105}{032602}{2014}.

\bibitem{Prigozhin10}
V. Sokolovsky, L. Prigozhin, and V. Dikovsky,  Supercond. Sci. Technol. {\bf 23}, 065003 (2010).

\bibitem{Brandt00}
E. H. Brandt, and G. Mikitik,  \PRL{85}{4164}{2000}.

\bibitem{Sanchez01}
A. Sanchez, and C. Navau, 
\PRB{64}{214506}{2001}.
\bibitem{Chen08}
C. Navau, A. Sanchez, N. Del-Valle, and D.-X. Chen, J. Appl. Phys. {\bf 103}, 113907 (2008).
\bibitem{Guillem13}
G. Via, C. Navau, and A. Sanchez,  J. Appl. Phys. {\bf 113}, 093905 (2013).

\bibitem{Plourde}
B. L. T. Plourde, D. J. Van Harlingen, D. Yu. Vodolazov, R. Besseling, M. B. S. Hesselberth, and P. H. Kes,  \PRB{64}{145031}{2001}.

\bibitem{Gehm1998}
M.~E.~Gehm, K.~M.~O'Hara, T.~A.~Savard, and J.~E.~Thomas, \PRA{58}{3914}{1998}.

\bibitem{RomeroIsart2012}
O. Romero-Isart, L. Clemente, C. Navau, A. Sanchez, J. I. Cirac, \PRL{109}{147205}{2012}.


\end{thebibliography}

\begin{thebibliography}{27}


\expandafter\ifx\csname natexlab\endcsname\relax\def\natexlab#1{#1}\fi
\expandafter\ifx\csname bibnamefont\endcsname\relax
  \def\bibnamefont#1{#1}\fi
\expandafter\ifx\csname bibfnamefont\endcsname\relax
  \def\bibfnamefont#1{#1}\fi
\expandafter\ifx\csname citenamefont\endcsname\relax
  \def\citenamefont#1{#1}\fi
\expandafter\ifx\csname url\endcsname\relax
  \def\url#1{\texttt{#1}}\fi
\expandafter\ifx\csname urlprefix\endcsname\relax\def\urlprefix{URL }\fi
\providecommand{\bibinfo}[2]{#2}
\providecommand{\eprint}[2][]{\url{#2}}

\bibitem{VodolazovMaksimov}
D.Yu. Vodolazov, and I.L. Maksimov, Physica C {\bf 349}, 125-138 (2001).


\bibitem{BrandtIndenbomSM}
E. H. Brandt, and M. Indenbom,  \PRB{48}{12893}{1993}.

\bibitem{ZeldovClemSM}
E. Zeldov, J. R. Clem, M. McElfresh, and M. Darwin,  \PRB{49}{9802}{1994}.

\bibitem{PlourdeSM}
B. L. T. Plourde, D. J. Van Harlingen, D. Yu. Vodolazov, R. Besseling, M. B. S. Hesselberth, and P. H. Kes,  \PRB{64}{014503}{2001}.

\bibitem{Gehm1998SM}
M.~E.~Gehm, K.~M.~O'Hara, T.~A.~Savard, and J.~E.~Thomas, \PRA{58}{3914}{1998}.


%
%
%
%
%

%
%

\end{thebibliography}
\end{document}